# A rotary plasmonic nanoclock

Ling Xin[1], Chao Zhou[1], Xiaoyang Duan[1,2], and Na Liu[1,2*]

[1]Max Planck Institute for Intelligent Systems, Heisenbergstrasse 3, D-70569 Stuttgart, Germany

[2]Kirchhoff Institute for Physics, University of Heidelberg, Im Neuenheimer Feld 227, D-69120, Heidelberg, Germany

Email: na.liu@kip.uni-heidelberg.de

**One of the fundamental challenges in nanophotonics is to gain full control over nanoscale optical elements. The precise spatiotemporal arrangement determines their interactions and collective behavior. To this end, DNA nanotechnology is employed as an unprecedented tool to create nanophotonic devices with excellent spatial addressability and temporal programmability. However, most of the current DNA-assembled nanophotonic devices can only reconfigure among random or very few defined states. Here, we demonstrate a DNA-assembled rotary plasmonic nanoclock. In this system, a rotor gold nanorod can carry out directional and reversible 360° rotation with respect to a stator gold nanorod, transitioning among 16 well-defined configurations powered by DNA fuels. The full-turn rotation process is monitored by optical spectroscopy in real time. We further demonstrate autonomous rotation of the plasmonic nanoclock powered by DNAzyme-RNA interactions. Such assembly approaches pave a viable route towards advanced nanophotonic systems entirely from the bottom-up.**

Precise and reversible control of metal nanoparticles in space and time is a prerequisite to build dynamic plasmonic nanosystems with tailored optical functionality. However, realization of spatiotemporal control in three-dimensions with nanoscale accuracy poses great challenges. If one looks into the dynamic systems in nature, especially those on the molecular level, for instance, biological machines[1,2] in living cells, their functionality represents remarkable examples of high-fidelity spatiotemporal control in a complex setting. $F_oF_1$-adenosine triphosphate (ATP) synthase[3] is a rotary machine, which is fascinating particularly due to its sophisticated structure and extraordinary performance. It consists of interacting rotor and stator components that couple ATP synthesis/hydrolysis with a transmembrane proton translocation. By consuming chemical energy, its subunits can carry out astonishing 360° rotary motion with precise control over biological activity in space and time.

Despite the vigorous progress in dynamic plasmonics, rotary plasmonic nanosystems have been restricted to perform simple switching functions between on and off states.[4-8] Here, we demonstrate a DNA-assembled plasmonic nanoclock that can carry out 360° rotary motion, taking inspiration from ATP synthase. Figure 1a shows the schematic of the rotary device, which contains several functional units. The rotor is a gold nanorod (AuNR, 38 nm × 10 nm) assembled on a 10-helix DNA origami[9-11] bundle of 52 nm in length. Two 12-nt single-stranded foot strands (black) are extended from the two ends of the origami bundle. The lateral distance between the two feet is ~34.6 nm. The stator is another AuNR (38 nm × 10 nm) assembled on the bottom surface of a DNA origami plate (54 nm × 52 nm). Both the rotor and stator AuNRs are positioned on the respective origami surfaces via 10 staple extensions as binding sites. On the top surface of the plate, a ring-shaped origami structure[12] is immobilized through three pairs of adjacent scaffold crossovers with 6 unpaired bases (see Supplementary Figs. 1 and 2). 16 footholds are evenly arranged and extended from the ring-shaped origami, forming a circular track of ~33.2 nm in diameter. This dimension matches the lateral distance between the two feet on the rotor bundle (see Supplementary Fig. 2). Each foothold consists of a binding domain (9-nt, black) with identical sequences and a toehold domain (8-nt, colored) for endowing specific site information. To ensure the structural flexibility for rotation, the origami bundle and the origami plate are linked through their centers using two adjacent scaffold crossovers with 30 unpaired bases (see Supplementary Figs. 1 and 2).

Figure 1b shows the arrangement of the 16 footholds ($fh_i$, $fh_{i'}$, $i = 1 - 8$) in 8 pairs distributed around the ring track. They are evenly separated by a distance of ~ 6.5 nm or by an angle of $\pi/8$. The footholds are identical in each pair and illustrated using the same color. The angle formed between the rotor and stator AuNRs is defined as $\theta$. Figure 1c depicts the working principle of the stepwise rotation powered by DNA fuels. The two AuNRs are omitted in this figure. The rotation is based on a 'release and capture' mechanism upon addition of specifically designed blocking and removal strands[13] (see Supplementary Table 1). The blocking and removal strands for $fh_1$, $fh_{1'}$ – $fh_8$, $fh_{8'}$ are marked as $B_1$–$B_8$ and $R_1$–$R_8$, respectively. Each blocking strand consists of three domains: an 11 nt-top domain (colored), a 6 nt-middle domain (black), and an 8 nt-bottom domain (colored). The top domain serves as a toehold and the rest two domains function as a blocking segment, which is complementary to one specific pair of the footholds on the ring track. The corresponding removal and blocking strands are fully complementary to each other. As shown in Fig. 1c, initially the two feet are bound to $fh_1$ and $fh_{1'}$ (red). The rest of the footholds are deactivated by blocking strands (only $fh_2$ and $fh_{2'}$ are shown for simplicity). To impose a clockwise rotation step from position 1-1' to position 2-2', removal strands $R_2$ and blocking strands $B_1$ are added sequentially. Through toehold-mediated strand displacement reactions[14] (see Supplementary Fig. 3), $fh_2$ and $fh_{2'}$ are first activated, and $fh_1$ and $fh_{1'}$ are subsequently blocked. The two feet released from position 1-1' search next active binding sites and subsequently capture $fh_2$ and $fh_{2'}$. This gives rise to a $\pi/8$ clockwise rotation and the origami bundle is bound to position 2-2'. Upon addition of corresponding DNA fuels (Supplementary Tables 2–4), the origami bundle and therefore the rotor AuNR can reach any designated positions around the ring track, carrying out programmable clockwise or counterclockwise stepwise rotation with respect to the stator AuNR. Figure 1d presents the transmission electron microscopy (TEM) image of the DNA origami structures. The origami bundle and ring are clearly visible in the inset image. Figure 1e shows the TEM image of the plasmonic nanoclock devices (see also Supplementary Fig. 4).

The crossed AuNRs in the plasmonic nanoclock system form a chiral object[8,15,16]. At incidence of circularly polarized light, the interacting plasmons excited in the two AuNRs can give rise to circular dichroism (CD) responses[8,16-18]. As a result, the stepwise rotations of the rotor AuNR with respect to the stator AuNR can be transformed into dynamic CD spectral changes, establishing a correlated relation between nanoscale motion and optical information. Figure 2a presents the CD spectra of the plasmonic nanoclock recorded during a full-turn clockwise rotation process at intervals of $\pi/8$ in 16 distinct states using a Jasco-1500 CD spectrometer. Nominally, at $\theta = 0$ the plasmonic nanoclock is in an achiral state. When $\theta$ increases from 0 to $\pi$, the system first enters a left-handed (LH) zone and then a right-handed (RH) zone by crossing an achiral state at $\theta = \pi/2$ in between. When $\theta$ increases from $\pi$ to $2\pi$, the system subsequently reaches LH and RH zones again. The theoretical results of the CD intensity as a function of $\theta$ are presented in Supplementary Fig. 5. In Fig. 2b, the experimental CD intensities at 732 nm (see the black dots marked in Fig. 2a) at different $\theta$ are replotted. It is apparent that the CD intensity follows a sinusoidal-like profile as a function of $\theta$. The experimental CD results show an overall LH preference mainly because the centers of the AuNRs, the origami bundle, and the ring track are not perfectly coaxial (see Supplementary Fig. 2) and the ring track is not a perfect circle. Nevertheless, such slight imperfections offer a unique opportunity to optically differentiate the rotation states in the two nominally identical LH (or RH) domains. It is noteworthy that the experimental CD intensity at $\theta = 2\pi$ returns to its initial value at $\theta = 0$. This reveals that the plasmonic nanoclock structures have been successfully directed back after the full-turn rotation, demonstrating the high fidelity of the programmable process. The *in situ* time-course measurements to characterize the kinetics of the individual rotation steps can be found in Supplementary Fig. 6. To validate the rotation reversibility, the CD intensities recorded along two opposite rotation directions are compared. As shown in Fig. 2c, the system first carries out counterclockwise stepwise rotation (red) from 0 to $15\pi/8$ and is then driven back, undergoing clockwise stepwise rotation (black). The CD intensities at the corresponding rotation angles along the two opposite routes nearly coincide with one another, proving excellent reversibility of the stepwise rotation process.

Next, we further develop the rotary plasmonic nanoclock to an autonomous system. This represents one of the first steps towards autonomous nanoplasmonics. The schematic of the autonomous system is shown in Fig. 3a. In this case, the feet are two 8-17 DNAzyme[19] strands (green) extended from the two ends of the DNA origami bundle. Each DNAzyme strand contains a 4-nt flexible segment, 7-nt and 9-nt binding domains, as well as a 14-nt catalytic core. Around the ring track on the origami plate, identical RNA strands (purple) are assembled, serving as footholds. The DNAzyme can catalyze the cleavage of the phosphodiester bond (purple dot) in the individual RNA substrate in the presence of $Mg^{2+}$, converting chemical energy into mechanical motion.[20-22] This gives rise to autonomous rotation of the rotor AuNR with respect to the stator AuNR processively. To specifically define the starting position of the autonomous process, two upper locking strands (black) are assembled on the origami bundle adjacent to the two DNAzyme feet, respectively, as shown in Fig. 3b and Supplementary Fig. 7. On the ring track, $fh_1$ and $fh_{1'}$ work as the corresponding lower locking strands (black-red) for hybridization with the upper ones to fix the origami bundle at position 1-1′ before rotation. Meanwhile, the two feet are initially deactivated using blocking strands b (brown, see Fig. 3b (i)). The TEM image is shown in Supplementary Fig. 8.

The working principle of the autonomous rotation without external intervention is depicted in Fig. 3b. Only the substrates at $fh_2$, $fh_{2'}$, $fh_3$, and $fh_{3'}$ are shown for simplicity. The autonomous rotation is actuated in two steps. The first one is the activation of the two DNAzyme feet by addition of removal strands r (Supplementary Fig. 9 I–III). The feet then bind to the neighboring substrates at position 2-2′, forming bulged pseudo-duplexes and subsequently cut the substrates into shorter segments. At this state, the rotor bundle resides likely in between positions 1-1′ and 2-2′. The second step is the releasing of the rotor bundle from position 1-1′ by addition of blocking strands $B_1$ (red-black-red) (Supplementary Fig. 9 IV–VI). Subsequently, the rotor AuNR hosted on the origami bundle starts autonomous movements powered by RNA hydrolysis (Supplementary Fig. 9 VII–X), processively rotating clockwise along the ring track based on a burnt-bridge mechanism[23].

To provide insights into the correlation between autonomous nanoscale motion and optical information as well as the fundamental differences between autonomous rotation without external intervention and stepwise rotation by sequential additions of DNA fuels, a set of *in situ* optical measurements have been carried out. Figure 4a shows the different time-course measurements for clockwise autonomous rotations, in which the number of the substrate footholds is successively varied. To facilitate clockwise rotations, position 8-8′ is free of substrates. The corresponding foothold positions of the six samples are presented in Supplementary Table 5. The CD intensities are monitored at a fixed wavelength of 732 nm. Upon addition of removal strands r, the CD intensities of the six samples exhibit similar signal increases. This indicates that the DNAzyme feet on the origami bundle are activated. The rotor AuNR starts to rotate and resides in between positions 1-1′ and 2-2′. Upon addition of blocking strands $B_1$, the CD intensity changes of the six samples become substantially different. For the track arrangements 1–2, 1–2–3, and 1–2–3–4, the CD intensities increase. This is similar to the trend observed for the stepwise rotation in Fig. 2b, in which the CD signal reaches the maximum value at position 3-3′ ($\theta = \pi/4$) in the first LH zone. For the track arrangements 1–2–3–4–5, 1–2–3–4–5–6, and 1–2–3–4–5–6–7, the CD intensities first show clear leaps and then immediate decreases. The observed leaps prove that a majority of the structures in the individual samples follow clockwise rotations. Otherwise, as shown in Fig. 2c a counterclockwise rotation would first lead to a valley feature instead of a leap. Interestingly, the CD decreasing magnitudes reveal a relation of 1–2–3–4–5–6–7 < 1–2–3–4–5–6 (see Fig. 4a). In other words, the minimum CD value occurs at position 6-6′ ($\theta = 5\pi/8$). It is different from the stepwise rotation in Fig. 2b, in which the CD signal reaches the minimum value at position 7-7′ ($\theta = 3\pi/4$). This counterintuitive observation outlines crucial differences of the plasmonic nanoclock system between carrying out autonomous rotation without external intervention and stepwise rotation powered by DNA fuels.

In order to understand the observed phenomena, two control experiments have been performed. We first examine whether the autonomous rotation can proceed, if two neighboring foothold sites along the ring track are free of substrates (Supplementary Table 6). As shown in Fig. 4b, for the track arrangement 1–2–x–x–5, *i.e.*, the substrates are omitted at $fh_3$ ($fh_{3'}$) and $fh_4$ ($fh_{4'}$), its time-course result (light blue) is very similar to that of the track arrangement 1-2 (black). However, for the track arrangement 1–2–3–4–5 (light grey), the CD intensity decreases over time during rotation. Such experimental evidence indicates that the autonomous rotation cannot proceed, if two neighboring substrate footholds are missing. The rotor AuNR thus halts, before the first open site. Next, we examine whether the autonomous rotation can progress, if only one substrate foothold is omitted along the ring track. As shown in Fig. 4c, for the track arrangement 1–2–3–x–5, *i.e.*, the substrate is omitted at $fh_4$ ($fh_{4'}$), its time-course result (dark blue) lies in between those of the track arrangements 1–2–3 (dark grey) and 1–2–3–4–5 (light grey). This reveals that it is possible for the rotor AuNR to hop over the open site and proceed with the autonomous rotation in this case.

Consequently, the observed phenomena in Fig. 4a can be understood and interpreted as follows. In the case of stepwise rotation powered by sequentially adding DNA fuels, each foothold on the ring track is site specific. The rotation direction is thus fully deterministic and unidirectional. In contrast, for autonomous rotation powered by RNA hydrolysis each substrate foothold on the ring track is identical. Once the DNAzyme feet are activated from starting position 1-1′, in principle both clockwise and counterclockwise rotations are allowed. For the track arrangement 1–2–3–4–5–6, the clockwise autonomous rotation is highly preferable, because the neighboring substrates at $fh_{7'}$ ($fh_7$) and $fh_{8'}$ ($fh_8$) for facilitating the counterclockwise rotation are omitted. For the track arrangement 1–2–3–4–5–6–7, as only one substrate is missing at $fh_8$ ($fh_{8'}$), it is possible for the counterclockwise rotation to take place by hopping over one open site. Therefore, in this case the time-course result contains contributions from both clockwise and counterclockwise rotations. As demonstrated in Figs. 2b and 2c, clockwise and counterclockwise rotations starting from position 1-1′ give rise to CD values of opposite signs. Therefore, in Fig. 4a the CD decreasing magnitudes reveal a relation of 1–2–3–4–5–6–7 < 1–2–3–4–5–6. Taking a further step, different time-course measurements for counterclockwise autonomous rotations are performed. In this case, position 2-2′ is free of substrates to facilitate counterclockwise rotations. As shown in Fig. 4d (see also Supplementary Table 7), the experimental results agree with the predictions well, further validating our interpretations.

DNA nanotechnology opens a unique avenue to build programmable nanophotonic devices with coordinated motion from the bottom-up.[4,13,24-27] The precise spatiotemporal control of nanoscale optical elements endows photonic nanoarchitectures with unprecedented dynamic complexity, which is beyond the scope of other state of the art nanotechniques.[28,29] Importantly, by exploiting the assembly power of DNA origami together with the patterning capabilities of top-down nanotechniques, such as electron-beam lithography,[30,31] many new opportunities may arise, setting a solid foundation for futuristic research topics. For instance, fully addressable and programmable metasurfaces could be achieved by patterning DNA-assembled rotary AuNRs in two-dimensional lattices to create artificial nanostructured interfaces for manipulating light properties including phase, amplitude, and polarization.[32,33] Such a metasurface could shape light wavefronts via geometric phase, *e.g*. Pancharatnam-Berry phase[32], by independently controlling the in-plane orientation of each rotary AuNR (see Supplementary Fig. 10). Furthermore, besides metal nanoparticles other nanoscale optical elements such as single emitters could also be integrated together. The combination of the DNA origami technique and on-chip nanolithography will particularly enable excellent control of light-matter interactions on the nanoscale in a fully programmable manner.[34] The relative distances and orientations between the AuNRs and single emitters could be dynamically tailored. This will substantially enhance and impedance match the outgoing radiation of the emitters to the far-field. In turn, light from the far-field can be confined in a subwavelength volume by the AuNRs, leading to strong modifications of the decay rate, emission direction, and quantum efficiency of the emitter[35]. Such hybrid platforms could, in the first instance, endow nanophotonics with full addressability and programmability on a single chip to create functional optical devices that utilize the respective strengths of both bottom-up and top-down nanotechniques.

## Methods

**Design and preparation of the DNA origami structures**

Single-stranded scaffold DNA (p8064) was purchased from tilibit nanosystems. Staple, blocking, and removal strands were purchased from Sigma Aldrich. Substrates were purchased from Integrated DNA Technologies, Inc. Agarose for electrophoresis and SYBR Gold nucleic acid stain were purchased from Life Technologies. Uranyl formate for negative TEM staining was purchased from Polysciences, Inc.

The DNA origami system was designed using caDNAno software[36]. It consisted of a 10-helix bundle, a 6-helix ring, and a single-layer plate arranged in a 'honeycomb' lattice, which were connected through scaffold crossovers (design and sequence details can be found in Supplementary Fig. 1 and Table 1). The DNA origami structures for stepwise plasmonic nanoclocks were assembled from 10 nM scaffold strands and 100 nM of each set of the staple strands (10-fold excess) in a $0.5 \times$ TE buffer with 12 mM $MgCl_2$ and 5 mM NaCl using a 20 h annealing program (85°C 5 min, 70°C–61°C −1°C/min, 60°C–51°C −1°C/1 h, 50°C–22°C −1°C/20 min, and 15 °C hold). The DNA origami structures were purified by 0.7% agarose gel electrophoresis in a $0.5 \times$ TBE buffer with 11 mM $MgCl_2$ for 3 h at 8 V/cm in a gel box immersed in an ice-water bath, and extracted by BioRad Freeze N Squeeze spin columns. The DNA origami structures for autonomous plasmonic nanoclocks were assembled and purified in a similar way. First, the mixture of the scaffold and staple strands without the foot strands were annealed to form the main structures in a $1 \times$ TA buffer with 12.5 mM $Mg(OAc)_2$, pH 7.6. Second, the blocked foot strands were added and the mixture was incubated at 30 °C for 24 h. Design and sequence details can be found in Supplementary Fig. 11, Tables 1 and 8. The autonomous DNA origami structures were purified by 0.7% agarose gel in a $1 \times$ TA buffer with 15 mM $Mg(OAc)_2$, pH 7.6 for 3 h at 8 V/cm in a gel box immersed in an ice-water bath (Supplementary Fig. 12).

**DNA functionalization of the AuNRs and assembly of the plasmonic nanoclocks**

AuNRs (38 nm in length and 10 nm in diameter) were purchased from Sigma-Aldrich (Cat no 716812). Functionalization of the AuNRs with thiolated DNA (5′ HS-$T_{12}$: HS-TTTTTTTTTTTT, purchased from Sigma-Aldrich) was carried out following a low pH procedure. First, thiolated DNA strands were incubated with TCEP [tris(2-carboxyethyl)phosphine] for 2 h. The ratio of DNA: TECP was 1:200. Second, the AuNRs (1 nM, 750 µL) were spun down and the supernatant was removed. The AuNRs were then mixed with thiolated DNA strands (250 µM 10 µL). 845 µL modification buffer ($0.59 \times$ TBE, 0.023% SDS, pH=3) was added. 10 µL 5 M NaCl was added every 10 min for 9 times. 50 µL 1 M NaOH was added subsequently to adjust the pH value to ~ 8, and the final concentration of NaCl reached 0.5 M. The AuNRs functionalized with DNA were then purified by centrifugation. Five times of centrifugations at a rate of 8,000×g for 30 min were carried out. Each time, the supernatant was carefully removed and the AuNRs were resuspended in a 0.5 × TBE buffer containing 0.02% of SDS. The supernatant was then removed and the remained AuNRs were mixed with the DNA origami structures at a ratio of 10:1. The mixtures were annealed with the procedure of 35°C–31°C −1°C/1 h, 30°C–27°C −1°C/4 h, and held at 26 °C. Gel electrophoresis was used to purify the plasmonic nanoclocks with the same condition for purification of the DNA origami structures (Supplementary Fig. 13).

**CD characterizations**

The CD spectra were measured using a Jasco-1500 CD Spectrometer with a Quartz SUPRASIL cuvette (path length, 10 mm). All measurements were carried out at 25 °C. The concentration of the blocking and removal strands was 250 μM.

For the individual CD spectrum measurements (Fig. 2a), one sample at position 1-1′ was divided into 17 copies. Each copy was 80 μL and the respective DNA fuels were added to drive the individual systems to their designated positions (Supplementary Table 2). To keep the concentration constant, equal volume of $H_2O$ was added, if no DNA strands were added. After each addition, the samples were incubated at 25 °C for about 1 h. The final concentrations of the AuNRs in all the systems were about 1 nM.

The time-course measurement for the stepwise rotation from position 5-5′ to position 5′-5 was carried out as follows. A 200 μL solution of the stepwise plasmonic nanoclock sample was used. The CD signal at 732 nm was monitored using the time-scan acquisition mode with a data pitch of 1 s. The respective blocking and removal strands were added to enable a programmed route (Supplementary Table 3). The initial concentration of the AuNRs was about 1.2 nM. After the process, the total volume increase was only about 9.6 μL (4.6%).

The time-course measurements for the autonomous rotation from position 1-1′ were carried out as follows. 100 μL solutions of different autonomous plasmonic nanoclock samples were measured. The CD signals at 732 nm were monitored using the time-scan acquisition mode with a data pitch of 1 s. The removal strands r and blocking strands $B_1$ were added to actuate the autonomous rotation (Supplementary Table 9). The initial concentration of AuNRs was 0.8 ~1.2 nM. After the process, the total volume increase was only about 1.5 μL (1.5%).

# Data Availability

All the data reported in this paper are available from the corresponding author upon request.

# Additional Information

## Acknowledgements

This project was supported by the Volkswagen foundation, and the European Research Council (ERC Dynamic Nano) grant. We thank Marion Kelsch for assistance with TEM. TEM images were collected at the Stuttgart Center for Electron Microscopy.

## Author Information

### Affiliations

Max Planck Institute for Intelligent Systems, Heisenbergstrasse 3, D-70569 Stuttgart, Germany

Ling Xin, Chao Zhou, Xiaoyang Duan & Na Liu

### Contributions

L.X., C.Z. and N.L. conceived concept. L.X. and C.Z. designed the DNA origami nanostructures and the working principles. L.X. performed the experiments. X.D. carried out the theoretical calculations and figure illustrations. L.X. and N.L. wrote the manuscript. All authors discussed the results, analyzed the data, and commented on the manuscript.

## Competing Interests

The authors declare no competing financial interests.

## Corresponding Author

Correspondence to Na Liu.

## References

1. Kinbara, K. & Aida, T. Toward intelligent molecular machines: directed motions of biological and artificial molecules and assemblies. *Chem. Rev.* **105**, 1377–1400 (2005).

2. van den Heuvel, M. G. & Dekker, C. Motor proteins at work for nanotechnology. *Science* **317**, 333–336 (2007).

3. Okuno, D., Iino, R. & Noji, H. Rotation and structure of FoF1-ATP synthase. *J. Biochem.* **149**, 655–664 (2011).

4. Kuzyk, A. *et al.* Reconfigurable 3D plasmonic metamolecules. *Nat. Mater.* **13**, 862–866 (2014).

5. Kuzyk, A. *et al.* A light-driven three-dimensional plasmonic nanosystem that translates molecular motion into reversible chiroptical function. *Nat. Commun.* **7**, 10591 (2016).

6. Kuzyk, A., Urban, M. J., Idili, A., Ricci, F. & Liu, N. Selective control of reconfigurable chiral plasmonic metamolecules. *Sci. Adv.* **3**, e1602803 (2017).

7. Zhou, C., Xin, L., Duan, X., Urban, M. J. & Liu, N. Dynamic Plasmonic System That Responds to Thermal and Aptamer-Target Regulations. *Nano Lett.* **18**, 7395–7399 (2018).

8. Zhou, C., Duan, X. & Liu, N. DNA-Nanotechnology-Enabled Chiral Plasmonics: From Static to Dynamic. *Acc. Chem. Res.* **50**, 2906–2914 (2017).

9. Rothemund, P. W. Folding DNA to create nanoscale shapes and patterns. *Nature* **440**, 297–302 (2006).

10. Douglas, S. M. *et al.* Self-assembly of DNA into nanoscale three-dimensional shapes. *Nature* **459**, 414–418 (2009).

11. Dietz, H., Douglas, S. M. & Shih, W. M. Folding DNA into twisted and curved nanoscale shapes. *Science* **325**, 725–730 (2009).

12. Yang, Y. *et al.* Self-assembly of size-controlled liposomes on DNA nanotemplates. *Nat. Chem.* **8**, 476–483 (2016).

13. Zhou, C., Duan, X. & Liu, N. A plasmonic nanorod that walks on DNA origami. *Nat. Commun.* **6**, 8102 (2015).

14. Zhang, D. Y. & Seelig, G. Dynamic DNA nanotechnology using strand-displacement reactions. *Nat. Chem.* **3**, 103–113 (2011).

15. Lan, X. *et al.* Bifacial DNA origami-directed discrete, three-dimensional, anisotropic plasmonic nanoarchitectures with tailored optical chirality. *J. Am. Chem. Soc.* **135**, 11441–11444 (2013).

16. Hentschel, M., Schäferling, M., Duan, X., Giessen, H. & Liu, N. Chiral plasmonics. *Sci. Adv.* **3**, e1602735 (2017).

17. Yin, X., Schäferling, M., Metzger, B. & Giessen, H. Interpreting chiral nanophotonic spectra: the plasmonic Born–Kuhn model. *Nano Lett.* **13**, 6238–6243 (2013).

18. Urban, M. J. *et al.* Chiral Plasmonic Nanostructures Enabled by Bottom-Up Approaches. *Annu. Rev. Phys. Chem.* **70**, 305–329 (2019).

19. Santoro, S. W. & Joyce, G. F. A general purpose RNA-cleaving DNA enzyme. *Proc. Natl. Acad. Sci. USA* **94**, 4262–4266 (1997).

20. Lund, K. *et al.* Molecular robots guided by prescriptive landscapes. *Nature* **465**, 206–210 (2010).

21. Cha, T.-G. *et al.* A synthetic DNA motor that transports nanoparticles along carbon nanotubes. *Nat. Nanotechnol.* **9**, 39–43 (2014).

22. Peng, H., Li, X.-F., Zhang, H. & Le, X. C. A microRNA-initiated DNAzyme motor operating in living cells. *Nat. Commun.* **8**, 14378 (2017).

23. Antal, T. & Krapivsky, P. L. "Burnt-bridge" mechanism of molecular motor motion. *Phys. Rev. E* **72**, 046104 (2005).

24. Wang, D. *et al.* A DNA Walker as a Fluorescence Signal Amplifier. *Nano Lett.* **17**, 5368–5374 (2017).

25. Kopperger, E. *et al.* A self-assembled nanoscale robotic arm controlled by electric fields. *Science* **359**, 296–301 (2018).

26. Urban, M. J. *et al.* Gold nanocrystal-mediated sliding of doublet DNA origami filaments. *Nat. Commun.* **9**, 1454 (2018).

27. Xin, L. *et al.* Watching a Single Fluorophore Molecule Walk into a Plasmonic Hotspot. *ACS Photonics* **6**, 985–993 (2019).

28. Liu, N. & Liedl, T. DNA-Assembled Advanced Plasmonic Architectures. *Chem. Rev.* **118**, 3032–3053 (2018).

29. Kuzyk, A., Jungmann, R., Acuna, G. P. & Liu, N. DNA Origami Route for Nanophotonics. *ACS Photonics* **5**, 1151–1163 (2018).

30. Scheible, M. B., Pardatscher, G., Kuzyk, A. & Simmel, F. C. Single molecule characterization of DNA binding and strand displacement reactions on lithographic DNA origami microarrays. *Nano Lett.* **14**, 1627–1633 (2014).

31. Kershner, R. J. *et al.* Placement and orientation of individual DNA shapes on lithographically patterned surfaces. *Nat. Nanotechnol.* **4**, 557–561 (2009).

32. Zheng, G. *et al.* Metasurface holograms reaching 80% efficiency. *Nat. Nanotechnol.* **10**, 308–312 (2015).

33. Chen, X. *et al.* Dual-polarity plasmonic metalens for visible light. *Nat. Commun.* **3**, 1198 (2012).

34. Gopinath, A., Miyazono, E., Faraon, A. & Rothemund, P. W. Engineering and mapping nanocavity emission via precision placement of DNA origami. *Nature* **535**, 401–405 (2016).

35. Novotny, L. & Van Hulst, N. Antennas for light. *Nat. Photonics* **5**, 83–90 (2011).

36. Douglas, S. M. *et al.* Rapid prototyping of 3D DNA-origami shapes with caDNAno. *Nucleic Acids Res.* **37**, 5001–5006 (2009).

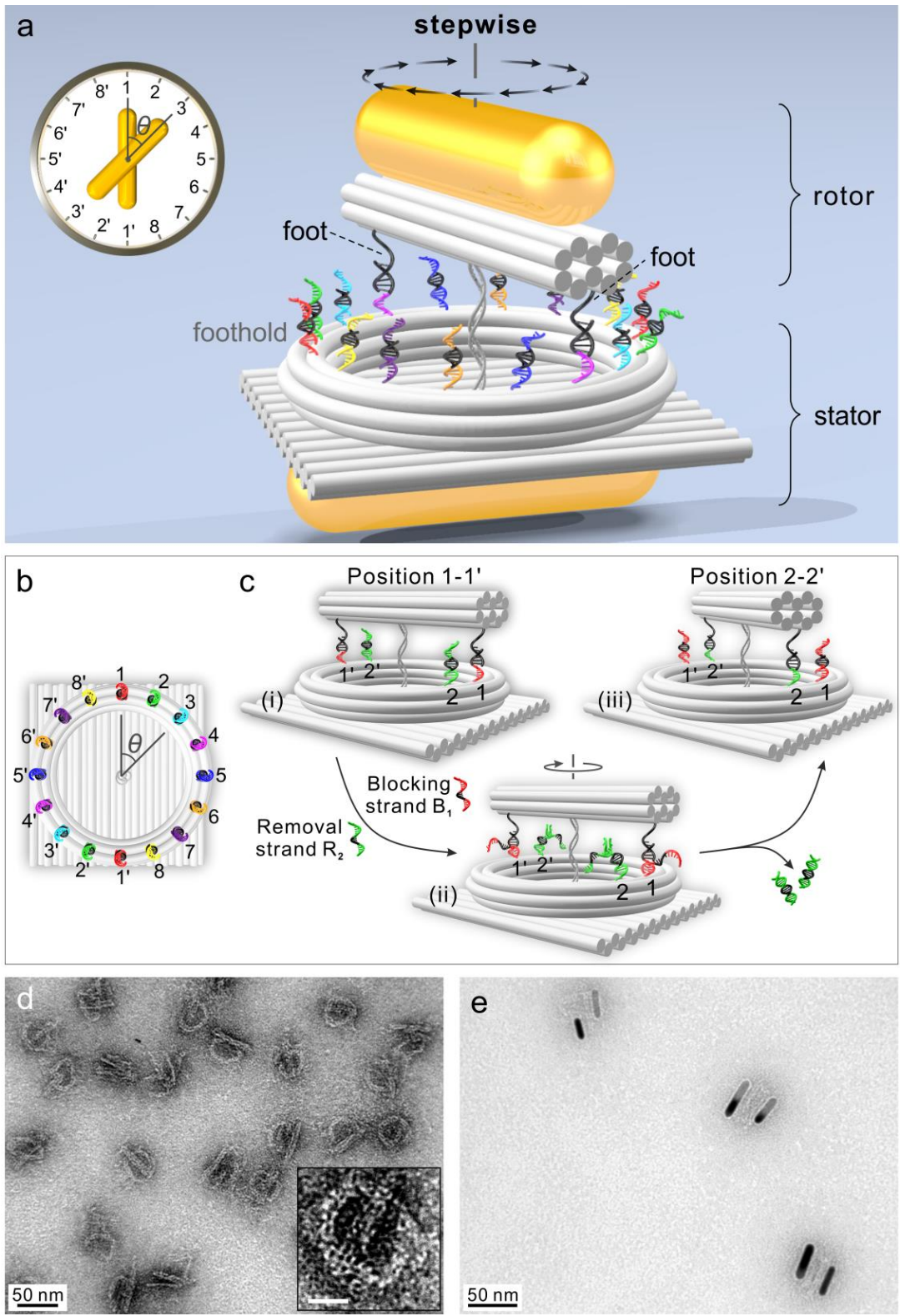

**Figure 1 | Stepwise plasmonic nanoclock. a**, Schematic of the rotary plasmonic nanoclock for stepwise rotation. The DNA origami structure consists of three connected

components: a bundle, a ring track, and a plate. One AuNR is assembled on the bundle, serving as rotor. The other AuNR is assembled on the back surface of the plate, serving as stator. 16 footholds in 8 pairs are evenly distributed around the origami ring, forming a circular track. The two feet (black) extended from the bundle can bind to any pair of the footholds. **b**, Foothold arrangement around the ring track. $fh_1$, $fh_{1'}$–$fh_8$, $fh_{8'}$ are shown in 8 different colors'. $\theta$ is defined as the angle between the rotor and stator AuNRs. Each rotation step corresponds to $\Delta\theta = \pi/8$. **c**, Working principle of the unidirectional stepwise rotation (AuNRs are not shown) powered by DNA fuels based on a 'release and capture' mechanism. It is enabled by addition of corresponding blocking strands and removal strands through toehold-mediated strand displacement reactions. **d,** TEM image of the DNA origami structures. Inset: enlarged view of a representative DNA structure, scale bar 20 nm. **e**, TEM image of the plasmonic nanoclocks.

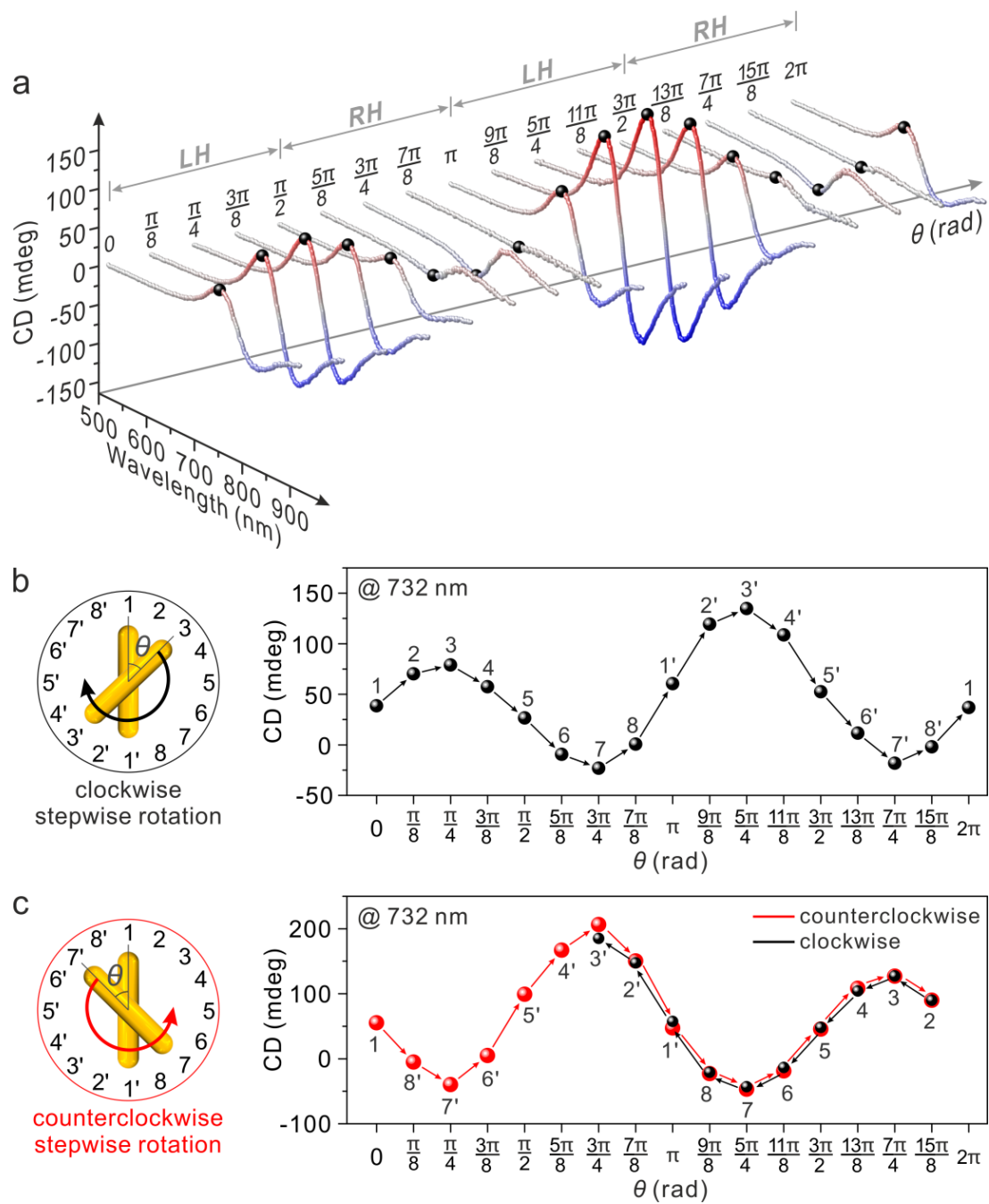

**Figure 2│Optical characterizations of the stepwise rotation process. a**, Circular dichroism spectra measured at various rotation angles $\theta$. **b**, CD intensities at 732 nm extracted from the spectra in Fig. 2a as a function of $\theta$ during a full-turn clockwise rotation process. **c**, Stepwise rotations along clockwise (black) and counterclockwise (red) directions, demonstrating good reversibility.

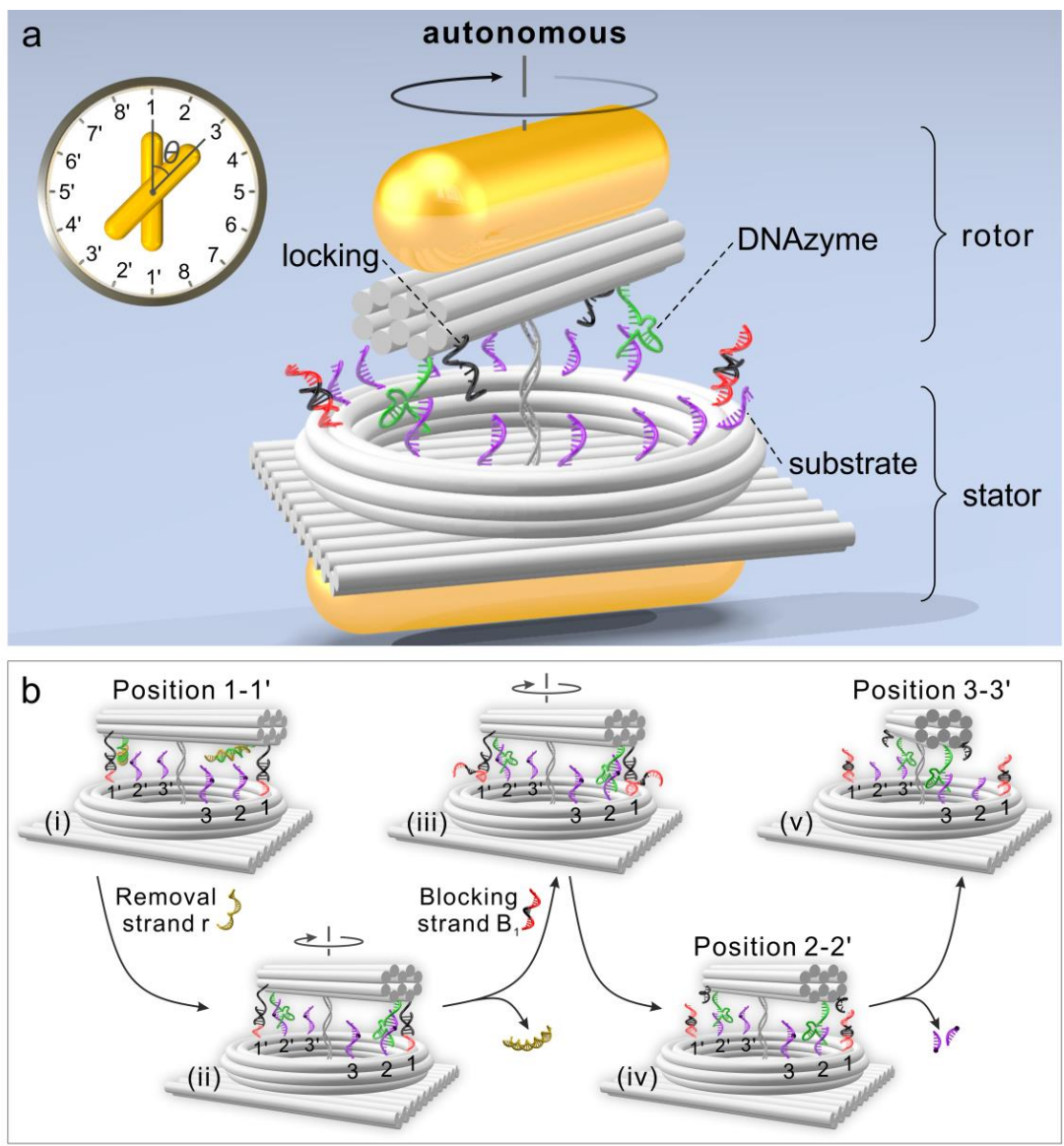

**Figure 3 | Autonomous plasmonic nanoclock. a**, Schematic of the rotary plasmonic nanoclock for autonomous rotation. The main DNA origami structure is taken from the stepwise plasmonic nanoclock design with modifications. The feet are two 8-17 DNAzyme strands (green) extended from the two ends of the bundle. RNA substrates (purple) are assembled around the ring track as footholds. Upper (black) and lower (black-red) locking strands are extended from the bundle and $fh_1$, $fh_{1'}$, respectively, to fix the rotor at starting position 1-1′. **b**, Working principle of the autonomous rotation powered by DNAzyme-RNA interactions in the presence of $Mg^{2+}$.

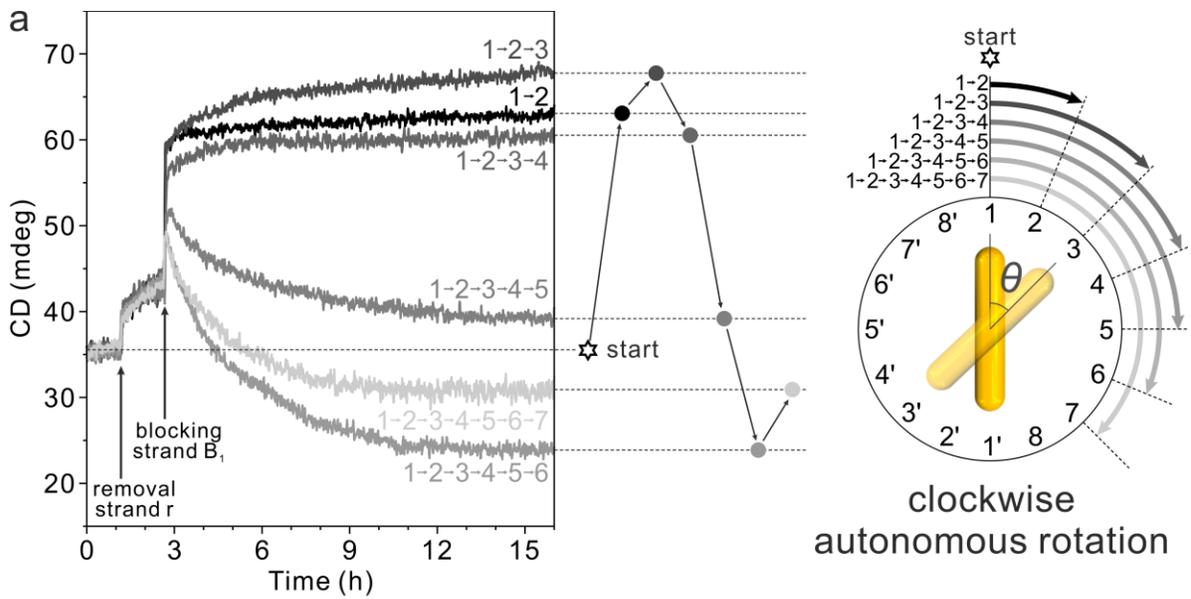
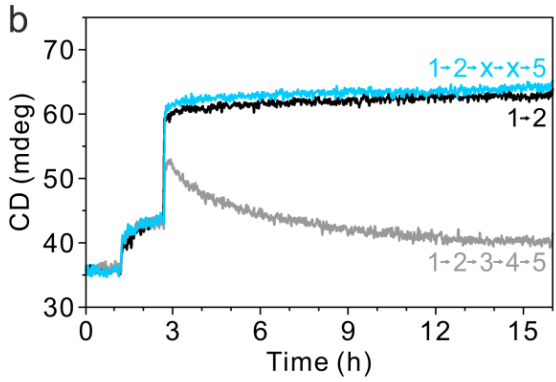
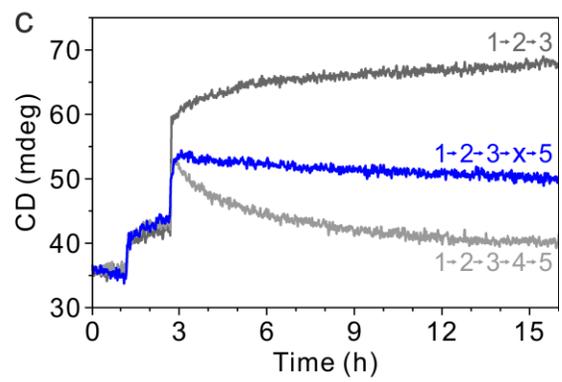
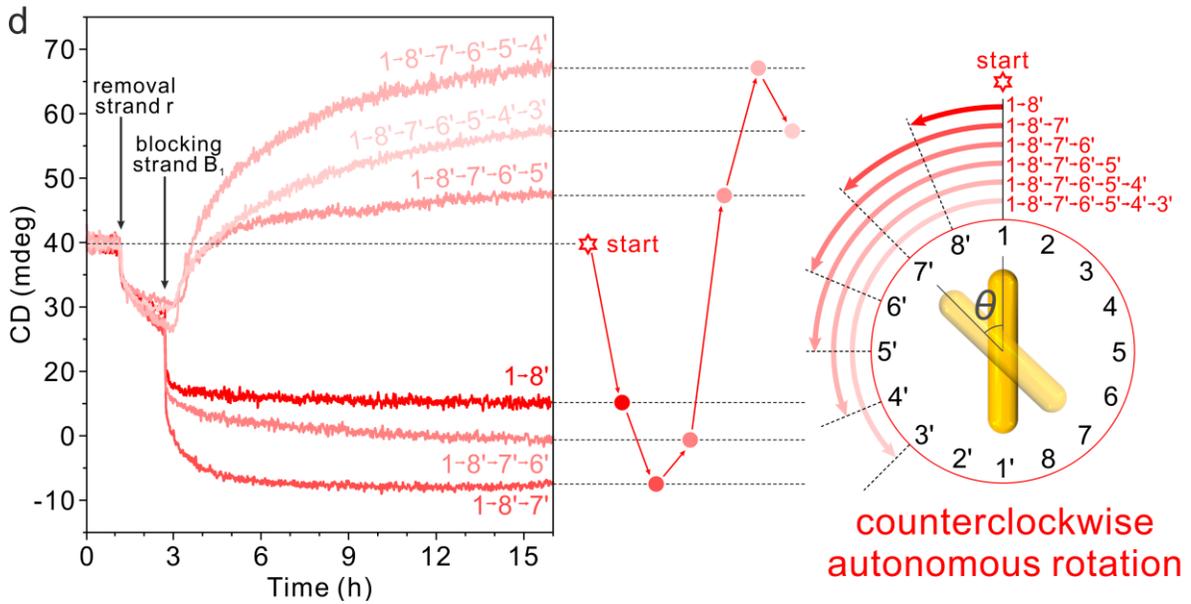

**Figure 4│*In situ* optical characterizations of the autonomous rotation process. a**, Time-course CD measurement at 732 nm for clockwise autonomous rotation. For each sample, the rotor is initially fixed at position 1-1′. The times for adding removal strands r to activate the DNAzyme feet and blocking strands $B_1$ to unlock the rotor from position 1-1′ are indicated in the plot. The arrangements of the substrates around the track are varied for different samples as illustrated in the schematic on the right (right half is shown). **b**, Control experiment 1, in which two substrates are omitted around the ring track. Track arrangement 1-2-x-x-5 (light blue) means that there are no substrates at $fh_3$ ($fh_{3'}$) and $fh_4$ ($fh_{4'}$). **c**, Control experiment 2, in which one substrate is omitted around the ring track. Track arrangement 1-2-3-x-5 (blue) means that there is no substrate at $fh_4$ ($fh_{4'}$). **d**, Time-course CD measurement at 732 nm for counterclockwise autonomous rotation. The arrangements of the substrates around the track are varied for different samples as illustrated in the schematic on the right (left half is shown).